\documentclass{article}
\usepackage{amsmath}
\usepackage{amssymb}
\usepackage{amsfonts}
\usepackage{bm}
\usepackage[dvips]{graphicx}

\begin{document}


%
\title{\begin{flushleft}
        {\large \underline{Article}}\vspace{\baselineskip}
        \end{flushleft}
{\bf Effect of Inertia on Linear Viscoelasticity \\
of Harmonic Dumbbell Model}}

\author{Takashi Uneyama\textsuperscript{1}\footnote{
E-mail: uneyama@mp.pse.nagoya-u.ac.jp},
Fumiaki Nakai\textsuperscript{2}, and Yuichi Masubuchi\textsuperscript{2}\\
\\
\textsuperscript{1} Center for Computational Science, Graduate School of Engineering, \\
Nagoya University, Furo-cho, Chikusa, Nagoya 464-8603, Japan\\
\textsuperscript{2} Department of Materials Physics, Graduate School of Engineering, \\
Nagoya University, Furo-cho, Chikusa, Nagoya 464-8603, Japan\\
}

\date{}

\maketitle



\begin{abstract}
 The overdamped (inertialess) dumbbell model is widely utilized to
 study rheological properties of polymers or other soft matters.
 In most cases, the effect of inertia is merely neglected because the
 momentum relaxation is much faster than the bond relaxation.
 We theoretically analyze the effect of inertia on the linear
 viscoelasticity of the harmonic dumbbell model. We show that
 the momentum and bond relaxation modes are kinetically coupled and the inertia
 can affect the bond relaxation if the momentum relaxation is not
 sufficiently fast. We derive an overdamped Langevin equation for the
 dumbbell model, which incorporates the weak inertia effect. Our model
 predicts the bond relaxation dynamics with the weak inertia effect
 correctly. We discuss how the weak inertia affects the linear
 viscoelasticity of a simple harmonic dumbbell model and the Rouse model.
\end{abstract}


\maketitle


\section{Introduction}
\label{introduction}

To study molecular-level dynamics of polymers or other soft matters,
simple Langevin equation models are useful. From the viewpoint of
rheology, the dynamics of molecular-level microscopic degrees of
freedom can be related to the macroscopic rheological functions such as
the shear relaxation modulus. For example, simple models such as the
Rouse model and the reptation model\cite{Doi-Edwards-book} can reproduce
experimentally obtained shear relaxation modulus data, and provide the
insights on the dynamics of polymers which cannot be directly
observed by microscopes.

In many cases, the overdamped Langevin equations without the inertia terms
are employed as the dynamic equations for polymers or soft matters.
This treatment is reasonable because the dynamical modes related to the inertia (momentum
relaxation modes) relax rapidly.
If we are interested in relatively long-time scales,
we can assume that the inertia effect is not important, and simply take
the overdamped (or inertialess) limit. For example, the mean-square
displacement (MSD) of a Brownian particle (which obeys the
Ornstein-Uhlenbeck process\cite{vanKampen-book}) shows ballistic behavior at the
short-time scale. But except the short-time scale region, the MSD can be
well approximated by a simple diffusion motion without inertia.
However, the effect of the inertia on the dynamics is generally not so
simple\cite{Sekimoto-1999,Matsuo-Sasa-2000}. The underdamped Langevin equations with the inertia terms are
the second order differential equations whereas the overdamped Langevin
equations are first order ones. The difference of the differential orders
qualitatively affects some properties of the model.
Therefore it is important to understand the effect of the inertia on the dynamics.

Here we limit ourselves to one of the simplest model, so-called the
dumbbell model\cite{Ottinger-book,Kroger-2004}.
Some researchers have investigated the effects of the inertia on
rheological
properties of the dumbbell model\cite{Booij-1988,Grassia-Hinch-1996,Schieber-Ottinger-1988,Degond-Liu-2009,Wuttke-2011}.
Roughly speaking, the overdamped Langevin equations are justified when
the inertia effect is very weak. Fortunately, this condition is satisfied for
many of polymeric solutions\cite{Grassia-Hinch-1996,Wuttke-2011}, and
thus the overdamped Langevin equations can be safely utilized to
describe the dynamics.
However, this does not mean that the overdamped Langevin equations are
always justified. For example, if we model the
relaxation process of relatively small objects such as segments, or if
we consider the higher order Rouse modes (which essentially behave as
statistically independent multiple dumbbell models), the inertia effect
may not be simply neglected.

In this work, we investigate the effect of the inertia on the linear
viscoelasticity of a simple dumbbell type model. We first derive the
approximate Langevin equation for the dumbbell model which is valid for
the weak (but non-zero) inertia case. Starting from the underdamped
Langevin equation with the inertia term, we show that an overdamped
Langevin equation with the lowest order correction by the inertia can be
obtained. The dynamics is slightly
accelerated by the inertia. Then we apply the obtained
approximate dynamic equation to the harmonic dumbbell for which we can
obtain analytic solutions. We calculate the shear relaxation modulus
and the longest relaxation time, with and without approximations, and
discuss how the inertia affects
the linear viscoelasticity. We also discuss how the inertia effect
accelerates the Rouse model, and a similar acceleration effect by the
memory effect in the generalized Langevin equation.

\section{Theory}
\label{theory}

\subsection{Model}
\label{model}

We start from a dumbbell model\cite{Ottinger-book,Kroger-2004} with an arbitrary bond potential and the
inertial terms.
We assume that a dumbbell consists of two beads with the same mass, $M$.
We describe the positions and momentum of two beads
as $\bm{R}_{1}$ and $\bm{R}_{2}$, and
$\bm{P}_{1}$ and $\bm{P}_{2}$, respectively. The Hamiltonian is
\begin{equation}
 \label{hamiltonian_dumbbell}
 \mathcal{H} = \frac{1}{2 M} (\bm{P}_{1}^{2} + \bm{P}_{2}^{2}) +
  U(\bm{R}_{1} - \bm{R}_{2}),
\end{equation}
where $U(\bm{R}_{1} - \bm{R}_{2})$ is the bond
potential (which is a harmonic function in the case of the harmonic
dumbbell model).
If we employ the Langevin thermostat to control the temperature of the system,
the dynamic equations for the dumbbell is
given as:
\begin{align}
 \label{dynamic_equation_r_beads}
 \frac{d\bm{R}_{i}(t)}{dt} & = \frac{1}{M} \bm{P}_{i}(t) , \\
 \label{dynamic_equation_p_beads}
 \frac{d\bm{P}_{i}(t)}{dt} & = - \frac{\partial
 U(\bm{R}_{1} - \bm{R}_{2}(t))}{\partial \bm{R}_{i}(t)} - \frac{Z}{M}
 \bm{P}_{i}(t) + \sqrt{2 Z k_{B} T} \bm{W}_{i}(t) , 
\end{align}
for $i = 1$ and $2$. Here, $Z$ is the friction coefficient for a bead, $k_{B}$ is
the Boltzmann constant, $T$ is the temperature, and
$\bm{W}_{i}(t)$ is a Gaussian white noise. The first and second moments
of the noise is given as follows:
\begin{equation}
 \label{fluctuation_dissipation_beads}
 \langle \bm{W}_{i}(t) \rangle = 0, \qquad
 \langle \bm{W}_{i}(t) \bm{W}_{j}(t') \rangle = \delta_{ij} \bm{1}
 \delta(t - t'), \qquad
\end{equation}
where $\langle \dots \rangle$ represents the statistical average and
$\bm{1}$ is the unit tensor.

We may rewrite the dynamic equation by introducing the center of mass
$\bar{\bm{R}} \equiv (\bm{R}_{1} + \bm{R}_{2}) / 2$ and the bond vector $\bm{r} \equiv
\bm{R}_{2} - \bm{R}_{1}$. We describe the momenta for the center of mass
and the bond vector as $\bar{\bm{P}} \equiv \bm{P}_{1} + \bm{P}_{2}$ and
$\bm{p} \equiv (\bm{P}_{2} - \bm{P}_{1}) / 2$. The Hamiltonian
\eqref{hamiltonian_dumbbell} can be rewritten as
\begin{equation}
 \label{hamiltonian_dumbbell_modified}
 \mathcal{H} = \frac{1}{2 \bar{M}} \bar{\bm{P}}^{2} + \frac{1}{2 m}
  \bm{p}^{2} + U(\bm{r}),
\end{equation}
with the total mass $\bar{M} \equiv 2 M$ and the reduced mass $m \equiv M / 2$.
Then we have
statistically independent two set of dynamic equations:
\begin{align}
 \label{dynamic_equation_r_cm}
 \frac{d\bar{\bm{R}}(t)}{dt} & = \frac{1}{\bar{M}} \bar{\bm{P}}(t), \\
 \label{dynamic_equation_p_cm}
 \frac{d\bar{\bm{P}}(t)}{dt} & = - \frac{\bar{Z}}{\bar{M}} \bar{\bm{P}}(t)
 + \sqrt{2  \bar{Z} k_{B} T} \bar{\bm{W}}(t), 
\end{align}
and
\begin{align}
 \label{dynamic_equation_r_bond}
 \frac{d\bm{r}(t)}{dt} & = \frac{1}{m} \bar{\bm{p}}(t), \\
 \label{dynamic_equation_p_bond}
 \frac{d\bm{p}(t)}{dt} & = - \frac{\partial U(\bm{r}(t))}{\partial \bm{r}(t)}
 - \frac{\zeta}{m} \bm{p}(t) + \sqrt{2 \zeta k_{B} T} \bm{w}(t),
\end{align}
where $\bar{Z} \equiv 2 Z$,
$\zeta = Z / 2$, $\bar{\bm{W}}(t) \equiv (\bm{W}_{1}(t) + \bm{W}_{2}(t))
/ \sqrt{2}$ and $\bm{w}(t) \equiv (\bm{W}_{2}(t) - \bm{W}_{1}(t)) /
\sqrt{2}$. Here, it should be noticed that $\bar{\bm{W}}(t)$ and
$\bm{w}(t)$ are statistically independent Gaussian white noises, and both of
them have the zero mean and unit variance (in the same way as eq~\eqref{fluctuation_dissipation_beads}).

At the overdamped limit, we set $d\bar{\bm{P}}/dt = 0$ and $d\bm{p}/dt$
= 0 in eqs~\eqref{dynamic_equation_p_cm} and \eqref{dynamic_equation_p_bond}. Then eqs
\eqref{dynamic_equation_r_cm}-\eqref{dynamic_equation_p_bond} reduce to
the well-known dynamic equations for the fully overdamped 
dumbbell model:
\begin{align}
 \label{dynamic_equation_overdamped_cm}
 \frac{d\bar{\bm{R}}(t)}{dt} & =  \sqrt{\frac{2 k_{B} T}{\bar{Z}}} \bar{\bm{W}}(t), \\
 \label{dynamic_equation_overdamped_bond}
 \frac{d\bm{r}(t)}{dt} & = - \frac{1}{\zeta} \frac{\partial U(\bm{r}(t))}{\partial \bm{r}(t)}
 + \sqrt{\frac{2 k_{B} T}{\zeta}} \bm{w}(t) .
\end{align}
This approximation is usually justified due to the fact that the characteristic
relaxation time of the inertia, $\tau_{m} \equiv m / \zeta = \bar{M} / \bar{Z}$, is much faster than that of the bond (by
the competition between the restoring force and the friction force).
The
momentum relaxation time is common for the center of mass and the bond. 
Various properties including the linear viscoelasticity of eqs
\eqref{dynamic_equation_overdamped_cm} and \eqref{dynamic_equation_overdamped_bond}
have been studied, and can be found in literates\cite{Ottinger-book,Kroger-2004}.
The center of mass motion does not contribute to the linear
viscoelasticity, and the linear viscoelasticity is fully determined
by eq~\eqref{dynamic_equation_overdamped_bond}.
The properties of the overdamped dumbbell model with various bond potential
models such as the finitely extensible nonlinear elasticity (FENE)
model have been studied extensively\cite{Ottinger-book,Kroger-2004}.

\subsection{Approximation for Weak Inertia Case}
\label{approximation_for_weak_inertia_case}

To investigate the effect of the inertia, we consider an approximate
dynamic equation which is valid for small $\tau_{m} = m / \zeta$.
We rewrite eqs~\eqref{dynamic_equation_r_bond} and 
\eqref{dynamic_equation_p_bond} as the second order stochastic
differential equation for $\bm{r}$:
\begin{equation}
 \label{dynamic_equation_r_bond_modified}
 m \frac{d^{2}\bm{r}(t)}{dt^{2}} 
 = - \frac{\partial U(\bm{r}(t))}{\partial \bm{r}(t)}
 - \zeta \frac{d\bm{r}(t)}{dt} + \sqrt{2  \zeta k_{B} T} \bm{w}(t) .
\end{equation}
We want a first order stochastic
differential equation as an approximate form for eq~\eqref{dynamic_equation_r_bond_modified}.
For this purpose, we further rewrite eq~\eqref{dynamic_equation_r_bond_modified} as
\begin{equation}
 \label{dynamic_equation_r_bond_modified2}
 \frac{d\bm{r}(t)}{dt} 
 = \left(1 + \tau_{m} \frac{d}{dt} \right)^{-1} \left[ - \frac{1}{\zeta} \frac{\partial U(\bm{r}(t))}{\partial \bm{r}(t)}
 + \sqrt{\frac{2 k_{B} T}{\zeta}} \bm{w}(t) \right] ,
\end{equation}
where $(1 + \tau_{m} d /dt)^{-1}$ is the inverse operator of $(1 +
\tau_{m} d / dt)$, and is defined via the following relation for a given function $f(t)$:
\begin{equation}
 \left(1 + \tau_{m} \frac{d}{dt} \right)^{-1}
  \left[ \left(1 + \tau_{m} \frac{d}{dt} \right) f(t) \right] = f(t) .
\end{equation}
For small $\tau_{m}$, the operator can be expanded into the power series
of $\tau_{m}$, and thus we have a simple approximate form, $(1 +
\tau_{m} d/dt)^{-1} \approx 1 - \tau_{m} d/dt$. (We retain only the
first order term in $\tau_{m}$.) Then we have the approximate
form for the dynamic equation \eqref{dynamic_equation_r_bond_modified2} as
\begin{equation}
 \label{dynamic_equation_r_bond_modified_approx}
\begin{split}
  \frac{d\bm{r}(t)}{dt} 
 & \approx - \frac{1}{\zeta} \frac{\partial U(\bm{r}(t))}{\partial \bm{r}(t)}
 +  \sqrt{\frac{2 k_{B} T}{\zeta}} \left(1 + \tau_{m} \frac{d}{dt} \right)^{-1}\bm{w}(t) 
\\
 & \qquad + \frac{\tau_{m}}{\zeta} \left[ \bm{K}(\bm{r}(t)) \cdot \frac{d\bm{r}(t)}{dt} 
 + \frac{k_{B} T}{\zeta} \frac{\partial}{\partial
  \bm{r}(t)} \cdot \bm{K}(\bm{r}(t)) \right]  ,
\end{split}
\end{equation}
where 
$\bm{K}(\bm{r})$ is the potential curvature tensor defined
as
\begin{equation}
 \label{potential_curvature_tensor}
 \bm{K}(\bm{r}) \equiv \frac{\partial^{2}U(\bm{r})}
  {\partial\bm{r}\partial\bm{r}} ,
\end{equation}
and we have utilized the Ito formula\cite{Gardiner-book} to
calculate $(d / dt) \partial U(\bm{r}(t)) / \partial \bm{r}(t)$. In
eq~\eqref{dynamic_equation_r_bond_modified_approx}, we have terms which
contain $d\bm{r}(t)/dt$ in both sides. Rearranging these terms and we
have
\begin{equation}
 \label{dynamic_equation_r_bond_modified_approx2}
\begin{split}
  \frac{d\bm{r}(t)}{dt} 
 & \approx - \frac{\bm{\Lambda}(\bm{r}(t)) }{\zeta} \cdot \frac{\partial U(\bm{r}(t))}{\partial \bm{r}(t)}
 + \frac{k_{B} T}{\zeta} \frac{\partial}{\partial
  \bm{r}(t)} \cdot \bm{\Lambda}(\bm{r}(t))  \\
 & \qquad + \sqrt{\frac{2 k_{B} T}{\zeta}} \bm{\Lambda}(\bm{r}(t)) \cdot \left(1 + \tau_{m} \frac{d}{dt} \right)^{-1} \bm{w}(t) ,
\end{split}
\end{equation}
with the effective (dimensionless) mobility tensor defined as
\begin{equation}
 \label{mobility_tensor}
 \bm{\Lambda}(\bm{r}) \equiv \bm{1} + \frac{\tau_{m}}{\zeta}
  \bm{K}(\bm{r})
  = \bm{1} + \frac{\tau_{m}}{\zeta}
  \frac{\partial^{2} U(\bm{r})}{\partial \bm{r} \partial \bm{r}} .
\end{equation}
The third term in the right-hand side of
eq~\eqref{dynamic_equation_r_bond_modified_approx2} can be interpreted
as the random noise. This noise is not well-defined (due to the
fact that time derivative of the noise $\bm{w}(t)$ is not well-defined) and the
fluctuation-dissipation relation seems not to be satisfied. It would be
natural for us to require
the thermodynamic equilibrium state can be correctly realized by the approximate
dynamic equation. This requirement can be achieved by modifying the noise term in
eq~\eqref{dynamic_equation_r_bond_modified_approx2} to recover the
fluctuation-dissipation relation.
(Here, it should be noticed that such an ad-hoc modification cannot be
fully justified in general. For the case of the harmonic dumbbell,
however, the noise property does not directly affect the linear
viscoelasticity. Thus this approximation is not bad for the analyses in
what follows.)
Finally, we have the following overdamped
Langevin equation as an approximate dynamic equation:
\begin{equation}
 \label{dynamic_equation_r_bond_approx_final}
 \frac{d\bm{r}(t)}{dt} =
  - \frac{\bm{\Lambda}(\bm{r}(t)) }{\zeta} \cdot
  \frac{\partial U(\bm{r}(t))}{\partial \bm{r}(t)}
  + \frac{k_{B} T}{\zeta} \frac{\partial}{\partial \bm{r}(t)} \cdot
  \bm{\Lambda}(\bm{r}(t)) 
  + \sqrt{\frac{2 k_{B} T}{\zeta}} \bm{B}(\bm{r}(t)) \cdot \bm{w}(t) ,
\end{equation}
with $\bm{B}(\bm{r})$ being the noise coefficient matrix which satisfies
$\bm{B}(\bm{r}) \cdot \bm{B}^{\mathrm{T}}(\bm{r}) = \bm{\Lambda}(\bm{r})$.
Eq~\eqref{dynamic_equation_r_bond_approx_final} is the approximate dynamic
equation where the inertia effect is weak, but not absent. Even at the
long-time scale, eq~\eqref{dynamic_equation_r_bond_approx_final} does
not reduce to eq~\eqref{dynamic_equation_overdamped_bond}.
Only at the fully overdamped case where $\tau_{m} = 0$ (or $m = 0$), the mobility tensor
simply becomes the unit
tensor ($\bm{\Lambda}(\bm{r}) = \bm{1}$) and 
eq~\eqref{dynamic_equation_r_bond_approx_final} reduces to
eq~\eqref{dynamic_equation_overdamped_bond}.
The fact that the overdamped limit corresponds to the inertialess case
($m = 0$) is consistent with the result obtained by Schieber and \"{O}ttinger\cite{Schieber-Ottinger-1988}.

\subsection{Harmonic Dumbbell Model}
\label{harmonic_dumbbell_model}

Here we consider the linear viscoelasticity of the harmonic dumbbell
model with the inertia effect and study how the inertia affects the
linear viscoelasticity. We set $U(\bm{r}) = k \bm{r}^{2} / 2$ and assume
that the spring constant $k$ satisfies $4 k m < \zeta^{2}$. (This condition is
required to suppress the oscillatory behavior, which is out of the scope of the
current work.) At the fully overdamped case where $m = 0$, the shear
relaxation modulus simply reduces to the single-mode Maxwell model:
\begin{equation}
 \label{relaxation_modulus_harmonic_overdamped}
 G(t) = \nu k_{B} T e^{- 2 t / \tau_{b}} ,
\end{equation}
where $\nu$ is the number density of the dumbbell and $\tau_{b} \equiv \zeta
/ k$ is the bond relaxation time. In this subsection, we consider how
the shear relaxation modulus deviates from
eq~\eqref{relaxation_modulus_harmonic_overdamped} in the presence of the
inertia effect.

We start from the exact solution for the case of the underdamped
dumbbell model.
The stress of the single harmonic dumbbell can be expressed as the
Kramers form:
\begin{equation}
 \label{single_dumbbell_stress_harmonic}
\begin{split}
 \hat{\bm{\sigma}}(\lbrace \bm{R}_{i} \rbrace,\lbrace \bm{P}_{i} \rbrace)
 & \equiv k (\bm{R}_{2} - \bm{R}_{1}) (\bm{R}_{2} - \bm{R}_{1})
 - \frac{1}{M}(\bm{P}_{1}\bm{P}_{1} + \bm{P}_{2} \bm{P}_{2})
   \\
 & = k \bm{r} \bm{r} - \frac{1}{\bar{M}} \bar{\bm{P}} \bar{\bm{P}}
 - \frac{1}{m} \bm{p}\bm{p} .
\end{split}
\end{equation}
Then, the shear relaxation modulus is given by the auto-correlation
function of the stress tensor, by the Green-Kubo formula\cite{Evans-Morris-book}:
\begin{equation}
 \label{relaxation_modulus_green_kubo}
 G(t) = \frac{\nu}{k_{B} T}
  \left\langle \hat{\sigma}_{xy}(t)
   \hat{\sigma}_{xy}(0)
  \right\rangle .
\end{equation}
Here $\hat{\sigma}_{xy}(t)$ represents the $xy$ component of the stress
tensor \eqref{single_dumbbell_stress_harmonic} at time
$t$.
From eqs~\eqref{single_dumbbell_stress_harmonic} and \eqref{relaxation_modulus_green_kubo},
the shear relaxation modulus for the harmonic dumbbell can be expressed
in terms of correlation functions. Because $\bar{\bm{P}}$ is statistically independent of $\bm{r}$ and
$\bm{p}$, and the system is isotropic, the expression for the relaxation
modulus can be simplified as
\begin{equation}
 \label{relaxation_modulus_harmonic}
\begin{split}
   G(t) & = \frac{\nu}{k_{B} T} \bigg[ 
 \frac{1}{\bar{M}^{2}} \langle \bar{P}_{x}(t) \bar{P}_{x}(0) \rangle^{2}
 + \frac{1}{m^{2}} \langle p_{x}(t) p_{x}(0) \rangle^{2}
 + k^{2} \langle r_{x}(t) r_{x}(0) \rangle^{2} \\
 & \qquad - \frac{k}{m}
 \left[ \langle r_{x}(t) p_{x}(0) \rangle^{2}
  + \langle p_{x}(t) r_{x}(0) \rangle^{2}
  \right]
 \bigg] .
\end{split}
\end{equation}
Here we have utilized some relations for correlation functions such as $\langle r_{y}(t) r_{y}(0) \rangle =
\langle r_{x}(t) r_{x}(0) \rangle$ and $\langle r_{x}(t) r_{y}(0)
\rangle = 0$.

The correlation function for
$\bar{\bm{P}}$ can be easily obtained. From eq~\eqref{dynamic_equation_p_cm},
we have
\begin{equation}
 \label{correlation_function_p_p_cm}
 \langle \bar{\bm{P}}(t) \bar{\bm{P}}(0) \rangle = 
 \bar{M} k_{B} T e^{- t / \tau_{m} } \bm{1},
\end{equation}
with $\tau_{m} = \bar{M} / \bar{Z}$ being the momentum 
relaxation time. Also, the correlation functions
for $\bm{r}$ and $\bm{p}$ can be calculated from
eqs~\eqref{dynamic_equation_r_bond} and \eqref{dynamic_equation_p_bond}.
For the harmonic potential, the restoring force $- k \bm{r}$ is linear in $\bm{r}$,
and we can rewrite eqs~\eqref{dynamic_equation_r_bond} and
\eqref{dynamic_equation_p_bond} as
\begin{equation}
\label{dynamic_equation_r_p_bond_harmonic}
 \frac{d}{dt}
  \begin{bmatrix}
   \bm{r}(t) \\
   \bm{p}(t)
  \end{bmatrix}
  = 
  -
  \begin{bmatrix}
   0 & - 1 / m \\
   k & \zeta / m
  \end{bmatrix}
  \cdot
  \begin{bmatrix}
   \bm{r}(t) \\
   \bm{p}(t)
  \end{bmatrix}
  +
  \begin{bmatrix}
   0 \\
   \sqrt{2 \zeta k_{B} T}\bm{w}(t)
  \end{bmatrix} .
\end{equation}
Since eq~\eqref{dynamic_equation_r_p_bond_harmonic} is linear in
$[\bm{r},\bm{p}]^{\mathrm{T}}$, we can solve it by diagonalizing the
coefficient matrix. To calculate the linear viscoelasticity, we need the
two time correlation functions. The result is:
\begin{align}
 \label{correlation_function_r_r_bond_harmonic}
 \langle \bm{r}(t)\bm{r}(0) \rangle
 & = \frac{k_{B} T}{k (\lambda_{+} - \lambda_{-})}
 (\lambda_{+} e^{- \lambda_{-} t} - \lambda_{-} e^{- \lambda_{+} t})
 \bm{1} , \\
 \label{correlation_function_p_p_bond_harmonic}
 \langle \bm{p}(t)\bm{p}(0) \rangle
 & = \frac{m k_{B} T}{\lambda_{+} - \lambda_{-}}
 (\lambda_{+} e^{- \lambda_{+} t} - \lambda_{-} e^{- \lambda_{-} t})
 \bm{1} , \\
 \label{correlation_function_r_p_bond_harmonic}
 \langle \bm{r}(t)\bm{p}(0) \rangle
 & = - \langle \bm{p}(t)\bm{r}(0) \rangle
 = \frac{k_{B} T}{\lambda_{+} - \lambda_{-}}
 (e^{- \lambda_{-} t} - e^{- \lambda_{+} t}) \bm{1} ,
\end{align}
with the eigenvalues of the coefficient matrix defined as
\begin{equation}
 \label{eigenvalues_bond_harmonic}
 \lambda_{\pm} \equiv \frac{\zeta \pm \sqrt{\zeta^{2} - 4 k m}}{2 m} .
\end{equation}
See
Appendix~\ref{detailed_calculations_for_harmonic_dumbbell_model}
for the detailed calculations.
Eq~\eqref{eigenvalues_bond_harmonic} can be rewritten in terms of the
momentum relaxation time $\tau_{m} = m / \zeta$
and the bond relaxation time $\tau_{b} = \zeta / k$:
\begin{equation}
 \label{eigenvalues_bond_harmonic_modified}
 \lambda_{\pm} =
 \frac{1 \pm \sqrt{1- 4 \tau_{m} / \tau_{b} }}{2 \tau_{m}} ,
\end{equation}
and if two relaxation times are well separated, $\tau_{b} \gg \tau_{m}$,
we can approximate eigenvalues as
$ \lambda_{+} \approx 1 / \tau_{m}$ and
$ \lambda_{-} \approx 
(1 + \tau_{m} / \tau_{b}) / \tau_{b} $.

By substituting eqs~\eqref{correlation_function_p_p_cm} and 
\eqref{correlation_function_r_r_bond_harmonic}-\eqref{correlation_function_r_p_bond_harmonic}
into eq~\eqref{relaxation_modulus_harmonic}, we finally have the explicit
expression for the relaxation modulus:
\begin{equation}
 \label{relaxation_modulus_harmonic_final}
   G(t) =  \nu k_{B} T \left[ 
 e^{-2 t / \tau_{m}}
 + e^{- (1 + \sqrt{1 - 4 \tau_{m} / \tau_{b}}) \, t / \tau_{m}}
 + e^{- (1 - \sqrt{1 - 4 \tau_{m} / \tau_{b}}) \, t / \tau_{m}}
 \right] .
\end{equation}
The detailed calculations are shown in
Appendix~\ref{detailed_calculations_for_harmonic_dumbbell_model}.
From eq~\eqref{relaxation_modulus_harmonic_final}, the relaxation
modulus consists of three Maxwell models.
At the long-time region, only the longest relaxation mode in
eq~\eqref{relaxation_modulus_harmonic_final} survives. In addition,
if the momentum relaxation is sufficiently fast, $\tau_{b} \gg
\tau_{m}$, the relaxation time can be approximated as a simple form.
Thus we have the following approximate form:
\begin{equation}
 \label{relaxation_modulus_harmonic_approx}
   G(t) \approx \nu k_{B} T  e^{- 2 (1 + \tau_{m} / \tau_{b}) \, t / \tau_{b}} .
\end{equation}
Eq~\eqref{relaxation_modulus_harmonic_approx}
is similar to eq~\eqref{relaxation_modulus_harmonic_overdamped} but the
relaxation time is slightly different. The relaxation time of
eq~\eqref{relaxation_modulus_harmonic_approx} is
$\bar{\tau} \equiv \tau_{b} /
2 ( 1 + \tau_{m} / \tau_{b})$.
This relaxation time $\bar{\tau}$ coincides to $\tau_{b} / 2$ if $\tau_{m} / \tau_{b} = 0$, but it
is generally different from $\tau_{b} / 2$. Therefore, we conclude that the
overdamped limit cannot be justified unless $\tau_{m} / \tau_{b}$ is
sufficiently small.

Now we apply the approximate dynamic equation
\eqref{dynamic_equation_r_bond_approx_final} to the harmonic dumbbell case.
By substituting $U(\bm{r}) = k \bm{r}^{2} / 2$ to
eqs~\eqref{potential_curvature_tensor} and \eqref{mobility_tensor}, we have
\begin{equation}
 \label{potential_curvature_tensor_mobility_tensor_harmonic}
 \bm{K}(\bm{r}) = k \bm{1}, \qquad
  \bm{\Lambda}(\bm{r}) = \left( 1 + \frac{\tau_{m}}{\tau_{b}} \right) \bm{1} .
\end{equation}
From eq~\eqref{potential_curvature_tensor_mobility_tensor_harmonic}, we
find that for the harmonic dumbbell model, the mobility tensor is
independent of the bond vector $\bm{r}$.
Therefore, eq~\eqref{dynamic_equation_r_bond_approx_final} becomes the
following overdamped Langevin equation with the additive noise:
\begin{equation}
 \label{dynamic_equation_r_bond_approx_harmonic}
 \frac{d\bm{r}(t)}{dt} =
  - \frac{1}{\tau_{b}} \left(1 + \frac{\tau_{m}}{\tau_{b}} \right) \bm{r}(t)
  + \sqrt{\frac{2 k_{B} T}{\zeta}
  \left(1 + \frac{\tau_{m}}{\tau_{b}}\right)} \bm{w}(t) .
\end{equation}
Eq~\eqref{dynamic_equation_r_bond_approx_harmonic} is a simple linear
Langevin equation (or the Ornstein-Uhlenbeck process\cite{vanKampen-book}), and can be solved easily.
From eq~\eqref{dynamic_equation_r_bond_approx_harmonic}, we have the
correlation function as
\begin{equation}
 \label{correlation_function_r_r_bond_approx_harmonic}
  \langle \bm{r}(t) \bm{r}(0) \rangle
  = \frac{k_{B} T}{k} e^{-(1 + \tau_{m} / \tau_{b}) \, t / \tau_{b}} \bm{1} .
\end{equation}

The single dumbbell stress tensor is given as the partial average over the
momenta for eq~\eqref{single_dumbbell_stress_harmonic}:
\begin{equation}
 \label{single_dumbbell_stress_approx_harmonic}
 \hat{\bm{\sigma}}(\bm{r}) = k \bm{r} \bm{r} - 2 k_{B} T \bm{1} ,
\end{equation}
and the Green-Kubo formula \eqref{relaxation_modulus_green_kubo} gives
the following expression for the relaxation modulus:
\begin{equation}
 \label{relaxation_modulus_harmonic_approx2}
 G(t) = \frac{\nu k^{2}}{k_{B} T} \langle r_{x}(t) r_{x}(0) \rangle^{2} .
\end{equation}
Finally, from eqs~\eqref{correlation_function_r_r_bond_approx_harmonic} and
\eqref{relaxation_modulus_harmonic_approx2}, we have the following
explicit expression for the relaxation modulus:
\begin{equation}
 \label{relaxation_modulus_harmonic_approx2_final}
 G(t) = \nu k_{B} T e^{- 2 (1 + \tau_{m} / \tau_{b}) \, t / \tau_{b}} .
\end{equation}
Eq~\eqref{relaxation_modulus_harmonic_approx2_final} has the same form
as
eq~\eqref{relaxation_modulus_harmonic_approx}. Therefore, from the
approximate dynamic equation \eqref{dynamic_equation_r_bond_approx_final}, we
can successfully reproduce the relaxation time $\bar{\tau} = \tau_{b} /
2 ( 1 + \tau_{m} / \tau_{b})$. 

\section{Discussions}
\label{discussions}

\subsection{Relaxation of Harmonic Dumbbell Model}
\label{relaxation_of_harmonic_dumbbell_model}

The long-time dynamics of a Brownian particle with inertia can be well
approximated by that without inertia. The diffusion coefficients
calculated from the mean square displacement the long-time region, with
and without inertia, are exactly the same. Thus for a diffusion process,
we can reasonably approximate the underdamped Langevin equation by the
overdamped one at the long-time region where $t \gtrsim \tau_{m}$.
In contrast, the long-time dynamics of a dumbbell model with
inertia does not reduce to that without inertia.
The exact expression
for the relaxation modulus with inertia
(eq~\eqref{relaxation_modulus_harmonic_final}) has three relaxation
modes: The center of mass momentum relaxation model, the bond momentum
relaxation mode, and the bond relaxation mode. The degrees of freedom of
the center of mass ($\bar{\bm{R}}$ and $\bar{\bm{P}}$) give only a single mode,
which is essentially the same as the relaxation mode of a single Brownian particle.
On the other hand, the degrees of freedom of
the bond ($\bm{r}$ and $\bm{p}$) have two relaxation modes, unlike the
case of the center of mass. The relaxation times of these two modes
depend both on the momentum relaxation time $\tau_{m}$ and the bond
relaxation $\tau_{b}$. We may interpret that these two modes are
kinetically coupled. Thus even at the long-time region, the relaxation
time $\bar{\tau}$ depends on both $\tau_{b}$ and $\tau_{m}$.

\begin{figure}[h!]
 \centering
 {\includegraphics[width=0.7\linewidth,clip]{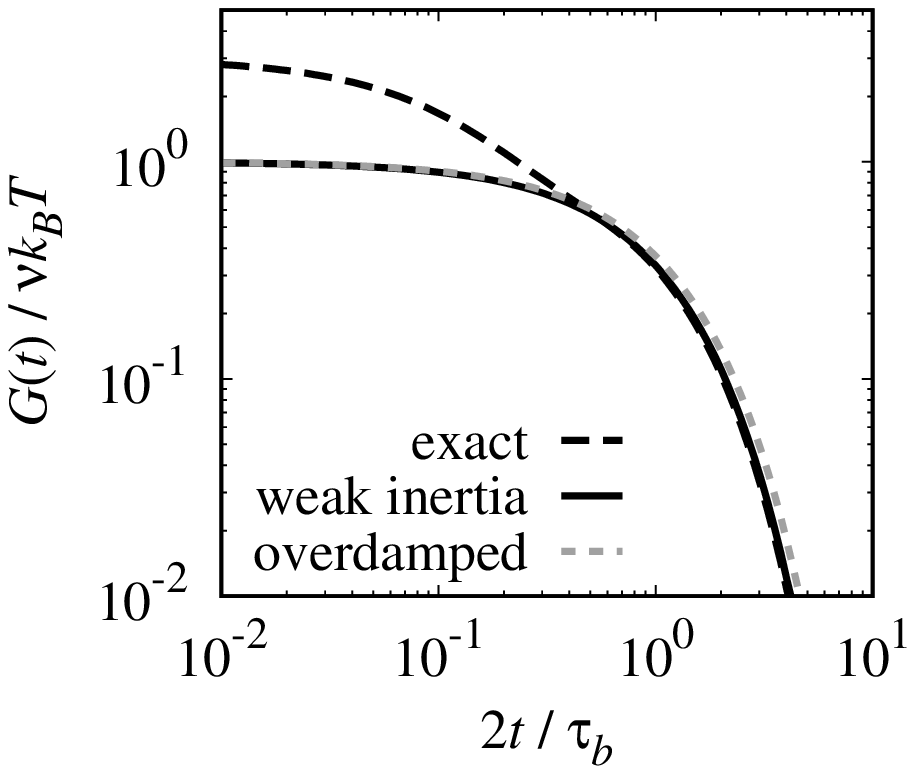}}
 \caption{}
 \label{dumbbell_model_relaxation_modulus}
\end{figure}

We show the relaxation modulus of the harmonic dumbbell with the inertia
effect ($\tau_{m} / \tau_{b} = 0.1$) in Figure~\ref{dumbbell_model_relaxation_modulus}.
At the short-time scale ($t \lesssim \tau_{m}$), we observe the momentum relaxation modes.
The modulus at the overdamped limit
(eq~\eqref{relaxation_modulus_harmonic_overdamped}) deviates from the
exact one (eq~\eqref{relaxation_modulus_harmonic_final}) even at the
long-time region ($t \gtrsim \tau_{b}$). (The deviation seems not large
at the logarithmic scale for $\tau_{m} / \tau_{b} = 0.1$.)
In contrast to the overdamped limit, the approximate form for weak inertia
(eq~\eqref{relaxation_modulus_harmonic_approx2_final}) agrees reasonably
with the exact one. Thus we find that the approximate dynamic equation
\eqref{dynamic_equation_r_bond_approx_final} derived in this work is a
reasonable approximation where the inertia effect is small but not fully negligible.

\begin{figure}[h!]
 \centering
 {\includegraphics[width=0.7\linewidth,clip]{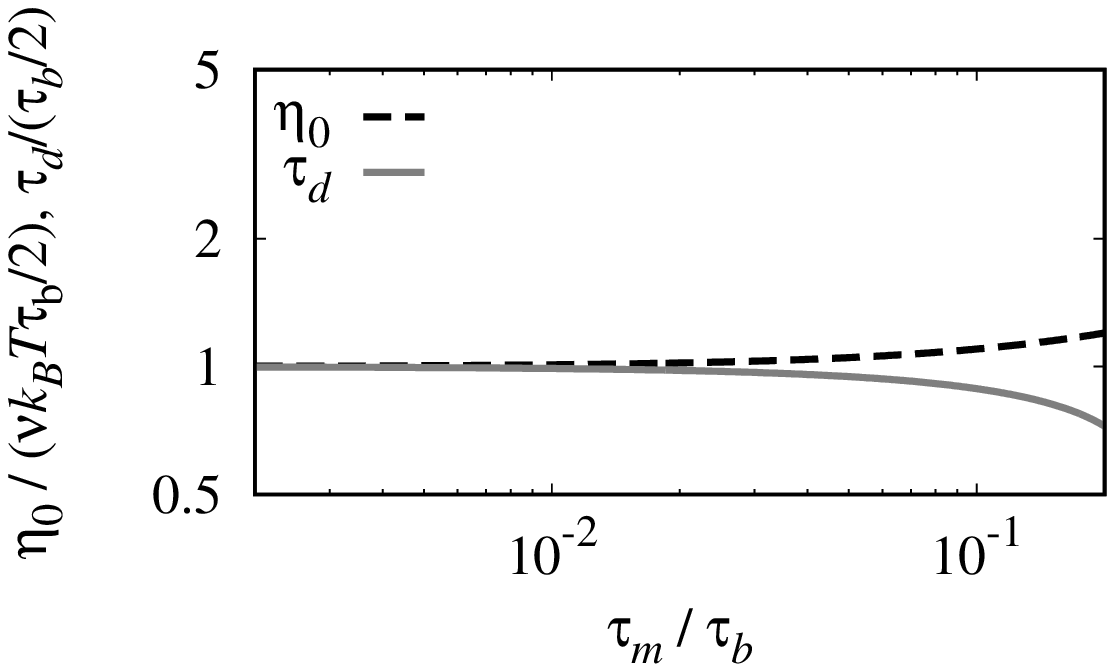}}
 \caption{}
 \label{dumbbell_model_viscosity_time}
\end{figure}

To study the effect of the inertia, we consider how the zero shear viscosity $\eta_{0}$ and
the longest relaxation time $\tau_{d}$ depend on the ratio $\tau_{m} / \tau_{b}$.
From eq~\eqref{relaxation_modulus_harmonic_final}, the longest relaxation time
is $\tau_{d} = \tau_{m} / (1 - \sqrt{1 - 4 \tau_{m} / \tau_{b}})$ (for $\tau_{m} / \tau_{b} < 1/4$).
The
zero shear viscosity at the overdamped limit is simply given as $\nu k_{B} T \tau_{b} / 2$.
On the other hand, from eq~\eqref{relaxation_modulus_harmonic_final}, the
zero shear viscosity with the inertia effet becomes
\begin{equation}
 \label{zero_shear_viscosity_harmonic}
  \frac{\eta_{0}}{\nu k_{B} T \tau_{b} / 2} = 
  \frac{2 \tau_{m}}{\tau_{b}}
  \left[ \frac{1}{2}
   + \frac{1}{1 + \sqrt{1 - 4 \tau_{m} / \tau_{b}}}
   + \frac{1}{1 - \sqrt{1 - 4 \tau_{m} / \tau_{b}}} \right].
\end{equation}
Figure~\ref{dumbbell_model_viscosity_time} shows the $\tau_{m}/\tau_{b}$ dependence
of the zero shear viscosity $\eta_{0}$ and the longset relaxation time $\tau_{d}$.
(For convenience, $\eta_{0}$ and $\tau_{d}$ are normalized by their overdamped limits.)
As $\tau_{m} / \tau_{b}$
increases, the zero shear viscosity increases wheareas the longest relaxation
time decreases. The increase of the viscosity is due to two relatively 
fast relaxation modes (the momentum relxation modes) which disappear at the overdamped limit.

We can understand the effect of the inertia term on the dynamics via the
Langevin equation \eqref{dynamic_equation_r_bond_approx_final}. In
eq~\eqref{dynamic_equation_r_bond_approx_final}, inertia term modulates the
the mobility tensor from $\bm{1}$ to
$\bm{\Lambda}(\bm{r})$. Intuitively, this can be interpreted as the
acceleration effect. The acceleration depends on the curvature tensor
$\bm{K}(\bm{r})$ which generally depends on the bond vector $\bm{r}$. In
the case of the harmonic potential, the curvature is isotropic and given
as a positive constant
$k$, and therefore all the dynamics is simply accelerated by the factor
$(1 + \tau_{m} / \tau_{b})$. This acceleration factor is consistent with
the fact that the ratio
of the relaxation time $\bar{\tau}$ to one at the overdamped limit
$\tau_{b} / 2 $ is $\bar{\tau} /
(\tau_{b} / 2) = (1 + \tau_{m} / \tau_{b})^{-1}$.
If the potential is not harmonic, the acceleration depends on the bond vector.
In such a case, the relaxation will be much more complex compared with
the harmonic dumbbell model. However, any bond potential potentials will
exhibit the acceleration effect on the bond dynamics, on average. Thus
we expect that the acceleration of the viscoelastic relaxation time will
be common and independent of the details of the bond potential.

\subsection{Rouse Model with Inertia Effect}
\label{rouse_model_with_inertia_effect}

The acceleration effect by the inertia term will be non-negligible if we
study relatively small scale relaxation modes. A simple yet nontrivial
example is the Rouse model\cite{Doi-Edwards-book}. Usually, the Rouse model is defined as the
the overdamped Langevin equation with the free energy for an ideal
chain which consists of $N$ beads. For simplicity, we assume that $N$ is
sufficiently large. We employ the underdamped
Langevin equation instead of the overdamped one:
\begin{equation}
 \label{dynamic_equation_rouse}
 m \frac{d^{2}\bm{R}_{k}(t)}{dt^{2}}
  = - \frac{\partial \mathcal{F}(\lbrace \bm{R}_{k}(t) \rbrace)}{\partial
  \bm{R}_{k}(t)} 
  - \zeta \frac{d\bm{R}_{k}(t)}{dt} + \sqrt{2 \zeta k_{B} T} \bm{w}_{k}(t) ,
\end{equation}
where $\bm{R}_{k}$ represents the position of the $k$-th bead ($k = 1, 2, \dots, N$), $m$ is the mass of a bead, $\zeta$
is the friction coefficient for a bead, $\mathcal{F}(\lbrace \bm{R}_{k} \rbrace)$ is the
free energy for an ideal chain, and $\bm{w}_{k}(t)$ is the Gaussian
white noise. The free energy is given as
\begin{equation}
 \mathcal{F}(\lbrace \bm{R}_{k} \rbrace) = \frac{3 k_{B} T}{2 b^{2}}
  \sum_{k = 1}^{N - 1} (\bm{R}_{k + 1} - \bm{R}_{k})^{2} ,
\end{equation}
with $b$ being the segment size, and the noise satisfies the following relations:
\begin{equation}
 \langle \bm{w}_{k}(t) \rangle = 0, \qquad
 \langle \bm{w}_{k}(t) \bm{w}_{l}(t') \rangle = \delta_{kl} \bm{1}
 \delta(t - t') .
\end{equation}
By using the Rouse mode defined as $\bm{X}_{p} \equiv \sqrt{2 / N} \sum_{k} \cos (p \pi k / N)
\bm{R}_{k}$ ($p = 1,2,3,\dots$), we can rewrite the dynamic equation
\eqref{dynamic_equation_rouse} as
\begin{equation}
 \label{dynamic_equation_rouse_modified}
 m \frac{d^{2} \bm{X}_{p}(t)}{d t^{2}}
  = - \frac{\zeta p^{2}}{\tau_{R}} \bm{X}_{p}(t)
 - \zeta \frac{d\bm{X}_{p}(t)}{dt}
 + \sqrt{2 \zeta k_{B} T} \bm{W}_{p}(t) ,
\end{equation}
where $\tau_{R} \equiv N^{2} b^{2} \zeta / 3 \pi^{2} k_{B} T$ is the
Rouse time and
$p = 1, 2, 3, \dots$ is the mode index (we ignore the zeroth mode $p = 0$,
because it does not contribute to the stress for the weak inertia
case). $\bm{W}_{p}(t)$ is the Gaussian white noise which satisfies
\begin{equation}
 \langle \bm{W}_{p}(t) \rangle = 0, \qquad
 \langle \bm{W}_{p}(t) \bm{W}_{q}(t') \rangle = \delta_{pq} \bm{1}
 \delta(t - t') . 
\end{equation}
We apply eq~\eqref{dynamic_equation_r_bond_approx_final} to the Rouse
model, eq~\eqref{dynamic_equation_rouse_modified}. Then, the approximate dynamic equation becomes
\begin{equation}
 \label{dynamic_equation_x_rouse_approx}
 \frac{d\bm{X}_{p}(t)}{dt}
  = - \frac{p^{2}}{\tau_{R}} 
  \left( 1 + \frac{\tau_{m} p^{2}}{\tau_{R}}
  \right) \bm{X}_{p}(t)
 + \sqrt{\frac{2 k_{B} T}{\zeta}
 \left( 1 + \frac{\tau_{m} p^{2}}{\tau_{R}}  \right)} \bm{W}_{p}(t) ,
\end{equation}
where $\tau_{m} = m / \zeta$ is the momentum relaxation time.
Eq~\eqref{dynamic_equation_x_rouse_approx} is valid only when the
inertia effect is weak. Roughly, this condition is estimated as $p \lesssim
p^{*}$, where $p^{*}$ is given as $p^{*}
= \sqrt{\tau_{R} / \tau_{m}}
= N b \zeta / (\pi \sqrt{3 m
k_{B} T})$.

The discussions above implies that we should introduce the cutoff for the
index $p$ unless the inertia effect is not fully negligible.
In molecular dynamics
simulations such as the Kremer-Grest model\cite{Kremer-Grest-1990}, we usually employ the unit
where the mass, energy, and the bead size becomes unity. Other
quantities such as the friction coefficient $\zeta$ and the temperature
$k_{B} T$ are
typically of the order of unity.
Then the maximum index $p^{*}$ can be estimated as $p^{*} \approx N / (\pi
\sqrt{3}) \approx 0.18 N$. Thus, we find that only less than $20\%$ of the Rouse modes
work as naively expected. (Higher order modes will largely deviate from the prediction
of the simple Rouse model.)

\begin{figure}[h!]
 \centering
 {\includegraphics[width=0.7\linewidth,clip]{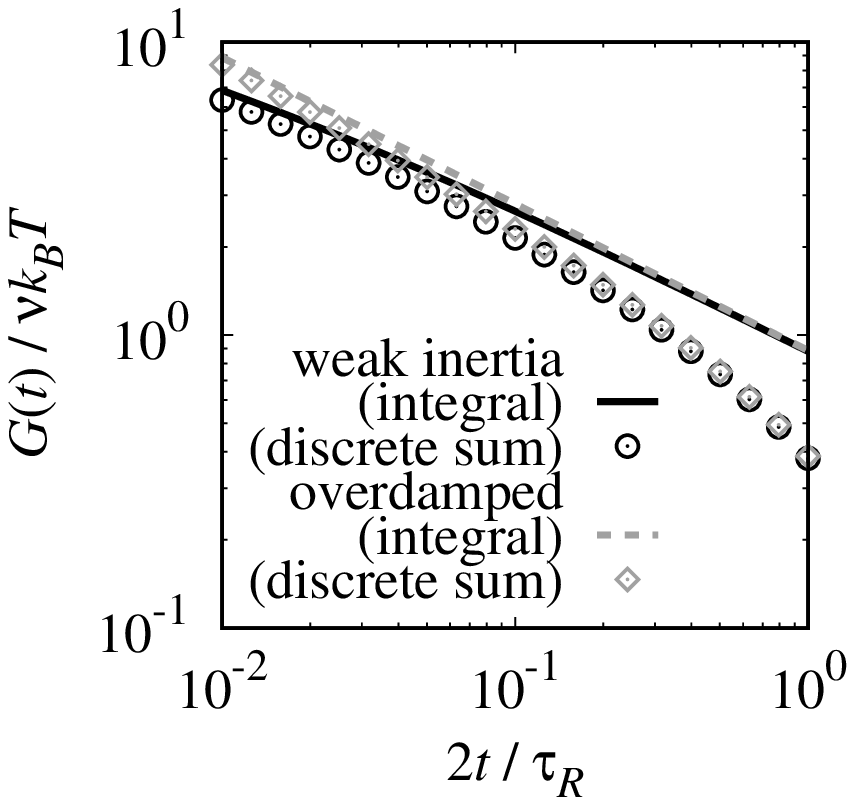}}
 \caption{}
 \label{rouse_model_relaxation_modulus}
\end{figure}

Even if the cutoff effect is minor and thus not important, the acceleration by the inertia
effect depends on the index $p$. The shear relaxation modulus can deviate from
the simple Rouse type behavior. From eq~\eqref{dynamic_equation_x_rouse_approx}, the relaxation modulus becomes
\begin{equation}
 \label{relaxation_modulus_rouse_exact}
 G(t) = \nu k_{B} T \sum_{p  = 1}^{ \infty} \exp
  \left[ - \frac{2 p^{2}}{\tau_{R}} 
  \left( 1 + \frac{\tau_{m} p^{2}}{\tau_{R}}
  \right) t \right] ,
\end{equation}
where $\nu$ is the chain density.
Therefore, the longest relaxation time is given as $\bar{\tau} = \tau_{R} / 2 (1 + \tau_{m}
/ \tau_{R})$. As the case of the harmonic dumbbell, the relaxation is
accelerated by the factor of $1 + \tau_{m} / \tau_{R}$.
For the intermediate time region where $\tau_{m} \lesssim t \lesssim \tau_{R}$, we may replace the discrete sum over $p$ by a continuum integral (as
the usual approximation for the Rouse model\cite{Doi-Edwards-book}), we have
\begin{equation}
 \label{relaxation_modulus_rouse_approx}
 \begin{split}
  G(t) & \approx \nu k_{B} T \int_{0}^{\infty} dp \, \exp
  \left[ - \frac{2 p^{2}}{\tau_{R}} 
  \left( 1 + \frac{\tau_{m} p^{2}}{\tau_{R}}
  \right) t \right] \\
  & = \frac{ \nu k_{B} T}{4} \sqrt{\frac{\tau_{R}}{\tau_{m}}} e^{t / 4 \tau_{m}}
  K_{1/4}
  \left( \frac{t}{4 \tau_{m}} \right) 
 \end{split}
\end{equation}
where $K_{1/4}(x)$ is the modified Bessel function of the second kind with the order $1/4$\cite{NIST-handbook}.
(The summation over $p$ for $p \ge 1$ in eq~\eqref{relaxation_modulus_rouse_exact}
is replaced by the integral for $p \ge 0$, not for $p \ge 1$, in eq~\eqref{relaxation_modulus_rouse_approx}.
The integral over the range $0 \le p \le 1$ is much smaller than that over the range $p \ge 1$, and thus
we simply neglected it.)
We can analytically calculate the complex modulus of the model
based on eq~\eqref{relaxation_modulus_rouse_exact} or
eq~\eqref{relaxation_modulus_rouse_approx}. The complex moduli given
by eqs~\eqref{relaxation_modulus_rouse_exact}
and \eqref{relaxation_modulus_rouse_approx} show reasonable agreement,
and thus we find that eq~\eqref{relaxation_modulus_rouse_approx} works as a
good approximation. The detailed calculations are shown in 
Appendix~\ref{complex_modulus_of_rouse_model}.
At relatively long-time scale ($\tau_{m} \ll t \lesssim \tau_{R}$), the asymptotic expansion\cite{NIST-handbook}
can be utilized and eq \eqref{relaxation_modulus_rouse_approx} can be
simplified as
\begin{equation}
 \label{relaxation_modulus_rouse_approx2}
  G(t) \approx \nu k_{B} T \sqrt{\frac{\pi \tau_{R}}{2 t}}
  \left( 1
  - \frac{3 \tau_{m}}{8 t} \right) .
\end{equation}
The first term in the parenthesis in the right-hand side of
eq~\eqref{relaxation_modulus_rouse_approx2} corresponds to the usual
(inertialess) Rouse type
terms\cite{Doi-Edwards-book}, and the second term can be interpreted as the correction.
Eqs~\eqref{relaxation_modulus_rouse_approx} and
\eqref{relaxation_modulus_rouse_approx2} imply that 
the shear relaxation modulus could slightly deviate from the well-known Rouse type behavior
$G(t) \propto t^{-1/2}$, if the inertia effect is
not sufficiently weak. 
We show the relaxation modulus by
eq~\eqref{relaxation_modulus_rouse_approx} (with $\tau_{m} / \tau_{R} =
0.01$) together with the usual
expression at the fully overdamped limit in
Figure~\ref{rouse_model_relaxation_modulus}.
For comparison, we show the relaxation modulus data directly
calculated with the discrete sum expression, eq~\eqref{relaxation_modulus_rouse_exact},
with $\tau_{m} / \tau_{R} = 0.01$ and $0$.
With this parameter
setting, the longest relaxation time $\bar{\tau}$ is almost the same as that at the
overdamped limit $\tau_{R} / 2$ ($\bar{\tau} / (\tau_{R} / 2) \approx
0.99$ for $\tau_{m} / \tau_{R} = 0.01$).
The relaxation modulus by eq~\eqref{relaxation_modulus_rouse_approx}
looks almost the same as that at the overdamped limit at the relatively
long-time region in Figure~\ref{rouse_model_relaxation_modulus}.
However, we observe a clear deviation between
eq~\eqref{relaxation_modulus_rouse_approx} and the overdamped limit 
at the relatively short-time region.
We observe a similar discrepancy for moduli calculated with
the discrete sum expression.
As the case of the dumbbell model, the inertia effect accelerates the relaxation.

Apparently, these results seem not to be consistent with the fact
that the experimental and simulation data of shear relaxation modulus for unentangled
polymer melts can be fitted well to the Rouse
model\cite{Likhtman-Sukumaran-Ramirez-2007,Masubuchi-Takata-Amamoto-Yamamoto-2018}. However, we should recall
that the polymer chains in melts are strongly interacting with each other by the
short-range repulsive interaction between segments, even if they are not entangled.
From this point of view, it is not plausible for unentangled polymer
melts to obey the simple Rouse model. Also, a recent study on the
Langevin equation with a fluctuating diffusivity suggested that the
Rouse type relaxation can emerge if the diffusivity (or the friction
coefficient) is heterogeneous and fluctuates in time\cite{Uneyama-Miyaguchi-Akimoto-2019}.
Therefore, even if the shear relaxation modulus exhibits the Rouse type
behavior, we cannot conclude that the molecular level dynamics fully obeys the
simple Rouse model. Careful and detailed analyses would be required to
further investigate the molecular level dynamics of polymers.

\subsection{Memory Effect}
\label{memory_effect}

To model complex relaxation behavior, the generalized Langevin equation\cite{Sekimoto-1999,Kawasaki-1973,Fox-1977}
is widely employed. In the generalized Langevin equation, the memory
effect is expressed by using a memory kernel. Formally, the generalized
Langevin equation is obtained by eliminating the microscopic
fast degrees of freedom, and the projection operator
formalism\cite{Kawasaki-1973} gives the expression of the
memory kernel.
Although, in principle, the memory kernel for a specific system can be
determined by the dynamics of the fast degrees of freedom in the system,
we cannot calculate its explicit form in most of practical cases.
Thus rather phenomoenolgical and simple memory functions such as a single
exponential form are employed in some cases.
In addition,
if the relaxation time of fast degrees of freedom, $\tau_{f}$, is much
faster than the characteristic time scale of the slow degrees of
freedom, the memory effect may be neglected. (This relaxation time
$\tau_{f}$ can be interpreted as the memory relaxation time.)
Then the usual Langevin equation can be used as an approximate dynamic
equation (the Markovian approximation).

In some aspects, the Markovian approximation explained above is similar to the
overdamped limit for the Langevin equation with the inertia effect.
Thus it would be informative to discuss how the memory kernel affects
the relaxation time of the dumbbell model.
Actually, the Langevin equation \eqref{dynamic_equation_r_bond_modified}
can be rewritten as the generalized Langevin equation, and thus we can
interpret our approximation method as an approximation for the
generalized Langevin equation.
We introduce the Green function $\Gamma(t)$ for the operator $(1 + \tau_{m}
d/dt)$:
\begin{equation}
 \label{kernel_function_definition}
 \left(1 + \tau_{m}\frac{d}{dt} \right) \Gamma(t) = \delta(t) .
\end{equation}
It is straightforward to see the following function $\Gamma(t)$
satisfies
eq~\eqref{kernel_function_definition} and thus is the Green function:
\begin{equation}
 \label{kernel_function}
  \Gamma(t) = \frac{1}{\tau_{m}} e^{-t / \tau_{m}} \Theta(t),
\end{equation}
where $\Theta(t)$ is the Heaviside step function. The application of the
operator $(1 + \tau_{m} d/dt)^{-1}$ to a given function $f(t)$ can be interpreted as
the convolution of $\Gamma(t)$ and $f(t)$:
\begin{equation}
 \left(1 + \tau_{m}\frac{d}{dt}\right)^{-1} f(t)
  = \int_{-\infty}^{t} dt' \, \Gamma(t - t') f(t') .
\end{equation}
Then eq~\eqref{dynamic_equation_r_bond_modified}
(or eq~\eqref{dynamic_equation_r_bond_modified2})
can be exactly rewritten as
\begin{equation}
 \label{dynamic_equation_r_bond_kernel}
 \frac{d\bm{r}(t)}{dt} 
 = - \frac{1}{\zeta} \int_{-\infty}^{t} dt' \, \Gamma(t - t') \frac{\partial U(\bm{r}(t'))}{\partial \bm{r}(t')}
 + \sqrt{\frac{k_{B} T}{\zeta}}\bm{\xi}(t) ,
\end{equation}
where $\bm{\xi}(t)$ is a Gaussian colored noise which is defined as
\begin{equation}
 \label{gaussian_colored_noise}
 \bm{\xi}(t) \equiv \sqrt{2 } \int_{-\infty}^{t} dt' \, \Gamma(t - t') 
 \bm{w}(t')  .
\end{equation}
Clearly, the first moment of $\bm{\xi}(t)$ is zero, $\langle \bm{\xi}(t)
\rangle = 0$. The second moment reduces to a simple form:
\begin{equation}
 \label{gaussian_colored_noise_second_order_moment}
  \begin{split}
   \langle \bm{\xi}(t) \bm{\xi}(t') \rangle
   & = 2 \int_{-\infty}^{t} ds \int_{-\infty}^{t'} ds' \, \Gamma(t - s) \Gamma(t' - s') 
   \bm{1} \delta(s - s') \\
   & = \frac{1}{\tau_{m}} \bm{1}
   e^{-|t - t'| / \tau_{m}} =  \bm{1} \Gamma(|t|) .
  \end{split}
\end{equation}
Eq~\eqref{gaussian_colored_noise_second_order_moment} is nothing but the
fluctuation-dissipation relation, and thus we can interpret
eq~\eqref{dynamic_equation_r_bond_kernel} as the generalized Langevin
equation with the memory kernel $\Gamma(t)$ given by eq~\eqref{kernel_function}.

As we discussed, the inertia effect accelerates the relaxation.
This acceleration effect is also caused in the case of the generalized
Langevin equation.
Therefore, in general, the memory effect may not be fully neglected even if
we consider relatively long-time scale dynamics.
To reproduce the correct relaxation behavior, we should employ the
effective mobility tensor.
We expect that 
the analyses for the weak inertia cases can be used to analyze the
dumbbell model with a weak memory effect, by replacing the 
momentum relaxation time $\tau_{m}$ by the memory relaxation time $\tau_{f}$.

\section{Conclusions}
\label{conclusions}

In this work, we studied the effect of the inertia on the dynamics of
the dumbbell model. We derived the approximate dynamic equation to
describe the dynamics for the weak inertia case. The inertia effects
modulate the mobility tensor, and thus the dynamics is slightly
accelerated. As a result, the dynamical properties such as the shear
relaxation modulus deviates from those at the overdamped limit.

As an example, we analyzed the harmonic dumbbell model which can be
solved analytically. We calculated the exact expression for the shear
relaxation modulus and showed that there are three relaxation modes.
One mode is the momentum relaxation of the center of mass, and the other two
modes are the momentum relaxation of the bond and the bond relaxation.
The relaxation times of two modes for the bond depend both on the
momentum relaxation time $\tau_{m}$ and the bond relaxation time
$\tau_{b}$. The longest relaxation time is approximated as $\bar{\tau}
\approx \tau_{b} / 2 (1 + \tau_{m} / \tau_{b})$, which is different from
that at the overdamped limit, $\tau_{b} / 2$. On the other hand, the approximate dynamic
equation \eqref{dynamic_equation_r_bond_approx_final} gives the correct long relaxation time.
Although the inertia effect may not be important in many practical
cases, it can be non-negligible for some limited cases. We consider that
the analyses based on the approximate dynamic equation and the detailed
and careful analyses of the experimental data will be required when the
inertia effect becomes non-negligible.

\section*{Acknowledgment}

The authors thank an anonymous reviewer to point that the 
sum in eq~\eqref{relaxation_modulus_rouse_exact} can be evaluated analytically
by utilizing the Laplace-Fourier transform.
This work was supported by Grant-in-Aid (KAKENHI) for Scientific
Research C 16K05513 and Grant-in-Aid (KAKENHI) for Scientific
Research A 17H01152.


\appendix

\section*{Appendix}

\section{Detailed Calculations for Harmonic Dumbbell Model}
\label{detailed_calculations_for_harmonic_dumbbell_model}

In this appendix, we show detailed calculations for the harmonic
dumbbell model with inertia effect, without approximations.
The dynamic equation is given as
eq~\eqref{dynamic_equation_r_p_bond_harmonic} in the main text.
We describe the coefficient matrix as
\begin{equation}
 \bm{C} \equiv 
  \begin{bmatrix}
   0 & - 1 / m \\
   k & \zeta / m
  \end{bmatrix} .
\end{equation}
$\bm{C}$ has the eigenvalues $\lambda_{\pm}$ in
eq~\eqref{eigenvalues_bond_harmonic} and the corresponding eigenvectors are
\begin{equation}
 \bm{\psi}_{\pm} \equiv
  \begin{bmatrix}
   1 \\
   - m \lambda_{\pm}
  \end{bmatrix} .
\end{equation}
Then we can diagonalize eq~\eqref{dynamic_equation_r_p_bond_harmonic}
with the transformation matrix $\bm{V} \equiv [\bm{\psi}_{+} \,
\bm{\psi}_{-}]$ and its inverse $\bm{V}^{-1}$. Multiplying $\bm{V}^{-1}$
from the left side to eq~\eqref{dynamic_equation_r_p_bond_harmonic} gives
\begin{equation}
\label{dynamic_equation_r_p_bond_harmonic_diagnoalized}
 \frac{d}{dt}
 \bm{V}^{-1} \cdot \begin{bmatrix}
   \bm{r}(t) \\
   \bm{p}(t)
 \end{bmatrix}
  = 
  -
  \begin{bmatrix}
   \lambda_{+} & 0 \\
   0 & \lambda_{-}
  \end{bmatrix} 
  \cdot \bm{V}^{-1} \cdot
  \begin{bmatrix}
   \bm{r}(t) \\
   \bm{p}(t)
  \end{bmatrix}
  + \bm{V}^{-1} \cdot
  \begin{bmatrix}
   0 \\
   \sqrt{2 \zeta k_{B} T}\bm{w}(t)
  \end{bmatrix} .
\end{equation}
After integrating
eq~\eqref{dynamic_equation_r_p_bond_harmonic_diagnoalized} from $0$ to
$t$, we multiply $\bm{V}$ to both sides and have
\begin{equation}
\label{dynamic_equation_r_p_bond_harmonic_solution}
\begin{split}
 \begin{bmatrix}
   \bm{r}(t) \\
   \bm{p}(t)
 \end{bmatrix}
 & =  \bm{V} \cdot 
  \begin{bmatrix}
   e^{-\lambda_{+} t} & 0 \\
   0 & e^{- \lambda_{-} t} 
  \end{bmatrix} \cdot \bm{V}^{-1} \cdot
  \begin{bmatrix}
   \bm{r}(0) \\
   \bm{p}(0)
  \end{bmatrix} \\
 & \qquad + \int_{0}^{t} dt' \,
  \bm{V} \cdot
  \begin{bmatrix}
   e^{-\lambda_{+} (t - t')} & 0 \\
   0 & e^{- \lambda_{-} (t - t')} 
  \end{bmatrix} \cdot
  \bm{V}^{-1} \cdot
  \begin{bmatrix}
   0 \\
   \sqrt{2 \zeta k_{B} T}\bm{w}(t')
  \end{bmatrix} .
\end{split}
\end{equation}
Now, $\bm{r}(0)$ and $\bm{p}(0)$ are statistically independent of the noise
$\bm{w}(t)$ for $t > 0$. In addition, in equilibrium, $\bm{r}(0)$ and
$\bm{p}(0)$ obey the Boltzmann distribution with the Hamiltonian \eqref{hamiltonian_dumbbell_modified}. Thus the averages over $\bm{r}(0), \bm{p}(0)$,
and $\bm{w}(t)$ can be took separately. The two time correlation functions for
$\bm{r}$ and $\bm{p}$ can be obtained as
\begin{equation}
\label{correlation_function_all_bond_harmonic}
 \begin{split}
\left\langle
  \begin{bmatrix}
   \bm{r}(t) \bm{r}(0) & \bm{r}(t) \bm{p}(0) \\
   \bm{p}(t) \bm{r}(0) & \bm{p}(t) \bm{p}(0) 
  \end{bmatrix}
  \right\rangle
  & =  \bm{V} \cdot 
  \begin{bmatrix} 
   e^{-\lambda_{+} t} & 0 \\
   0 & e^{- \lambda_{-} t} 
  \end{bmatrix} \cdot \bm{V}^{-1} \cdot
  \begin{bmatrix}
   (k_{B} T / k) \bm{1} & 0 \\
   0 & m k_{B} T \bm{1}
  \end{bmatrix} \\
  & = \frac{k_{B} T}{\lambda_{+} - \lambda_{-}}
  \begin{bmatrix}
   (\lambda_{+} e^{- \lambda_{-} t} - \lambda_{-} e^{-\lambda_{+} t} ) / k
  &  e^{- \lambda_{-} t} - e^{-\lambda_{+} t}  \\
   e^{- \lambda_{+} t} - e^{-\lambda_{-} t}
   &  m (\lambda_{+} e^{- \lambda_{+} t}- \lambda_{-} e^{- \lambda_{-}
   t} ) 
  \end{bmatrix}  \bm{1} ,
 \end{split}
\end{equation}
where we have utilized the relation for the eigenvalues, $\lambda_{+} \lambda_{-} = k /
m$. Eq~\eqref{correlation_function_all_bond_harmonic} gives
eqs~\eqref{correlation_function_r_r_bond_harmonic}-\eqref{correlation_function_r_p_bond_harmonic}
in the main text.

The relaxation modulus $G(t)$ can be calculated by
eqs~\eqref{relaxation_modulus_harmonic}, \eqref{correlation_function_p_p_cm} and 
\eqref{correlation_function_r_r_bond_harmonic}-\eqref{correlation_function_r_p_bond_harmonic}:
\begin{equation}
 \label{relaxation_modulus_harmonic_modified}
\begin{split}
   G(t) & = \nu k_{B} T \bigg[ 
 e^{-2 t / \tau_{m}}
 + \left(\frac{\lambda_{+} e^{- \lambda_{+} t} - \lambda_{-} e^{-
 \lambda_{-} t}}{\lambda_{+} - \lambda_{-}}\right)^{2}
 + \left(\frac{\lambda_{+} e^{- \lambda_{-} t} - \lambda_{-} e^{-
 \lambda_{+} t}}{\lambda_{+} - \lambda_{-}}\right)^{2} \\
 & \qquad - \frac{2}{\tau_{b} \tau_{m}}
 \left( \frac{e^{- \lambda_{-} t} - e^{- \lambda_{+} t}}{\lambda_{+} - \lambda_{-}} \right)^{2}
 \bigg] ,
\end{split}
\end{equation}
where we have utilized the relation $k / m = (k / \zeta) (\zeta / m) = 1
/ \tau_{b} \tau_{m}$.
Some factors in eq~\eqref{relaxation_modulus_harmonic_modified}
can be simplified as follows:
\begin{equation}
 \frac{1}{(\lambda_{+} - \lambda_{-})^{2}}
  = \frac{m^{2}}{\zeta^{2} - 4 k m}
  = \frac{\tau_{m}^{2}}{1 - 4 \tau_{m} / \tau_{b}} ,
\end{equation}
\begin{equation}
 \begin{split}
  & (\lambda_{+} e^{- \lambda_{+} t} - \lambda_{-} e^{- t \lambda_{-}})^{2}
  + (\lambda_{+} e^{- \lambda_{-} t} - \lambda_{-} e^{- t
  \lambda_{+}})^{2} \\
  & = \frac{1}{\tau_{m}^{2}}
  \left[
  \left( 1 - \frac{2 \tau_{m}}{\tau_{b}} \right) (e^{- 2 \lambda_{+} t} + e^{- 2 t \lambda_{-}})
  - \frac{4 \tau_{m}}{\tau_{b}} e^{- (\lambda_{+} + \lambda_{-}) t}
  \right] ,
 \end{split}
\end{equation}
\begin{equation}
   \frac{2}{\tau_{b} \tau_{m}}
 ( e^{- \lambda_{-} t} - e^{- \lambda_{+} t})^{2} \\
   = \frac{1}{\tau_{m}^{2}}
    \left[  \frac{2 \tau_{m}}{\tau_{b}} (e^{- 2 \lambda_{+} t} + e^{- 2 \lambda_{-} t})
  -  \frac{4 \tau_{m}}{\tau_{b}} e^{- (\lambda_{+} +\lambda_{-} ) t} \right] .
\end{equation}
Thus we have
\begin{equation}
 \label{relaxation_modulus_harmonic_modified2}
   G(t)
  = \nu k_{B} T \left( 
 e^{-2 t / \tau_{m}}
 + e^{- 2 \lambda_{+} t} + e^{- 2 \lambda_{-} t}
 \right) .
\end{equation}
Eq~\eqref{relaxation_modulus_harmonic_modified2} together with
eq~\eqref{eigenvalues_bond_harmonic}
gives eq~\eqref{relaxation_modulus_harmonic_final} in the main text.

\section{Complex Modulus of Rouse Model}
\label{complex_modulus_of_rouse_model}

In this appendix, we calculate the complex modulus $G^{*}(\omega)$ of the
Rouse model with inertia effect. The complex modulus is obtained as the
Fourier transform of the relaxation modulus $G(t)$. From the view point
of experiment, the complex modulus is much easier to measure and thus it
would be informative to show some explicit expressions for the complex modulus.

Firstly we start from eq~\eqref{relaxation_modulus_rouse_exact}. The complex
modulus $G^{*}(\omega)$ is calculated as
\begin{equation}
 \label{complex_modulus_rouse_exact}
 \begin{split}
  G^{*}(\omega) 
  & = i \omega \int_{0}^{\infty} dt \, G(t) e^{-i \omega t} \\
  & = \nu k_{B} T 
  \sum_{p = 1}^{\infty} 
  \frac{i \omega \tau_{R}^{2} / 2 \tau_{m}}{ p^{4} + (\tau_{R} / \tau_{m}) p^{2}
  + i \omega  \tau_{R}^{2} / 2 \tau_{m}} \\
  & = \nu k_{B} T 
  \sum_{p = 1}^{\infty} 
  \frac{\alpha_{+} \alpha_{-}}{ (p^{2} + \alpha_{+}) (p^{2} + \alpha_{-})} \\
  & = \nu k_{B} T 
  \frac{\alpha_{+} \alpha_{-}}{\alpha_{+} - \alpha_{-}}
  \sum_{p = 1}^{\infty} 
  \left[
  \frac{1}{p^{2} + \alpha_{-}} 
  - \frac{1}{p^{2} + \alpha_{+}} 
  \right] ,
 \end{split}
\end{equation}
where we have defined $\alpha_{\pm}$ as
\begin{equation}
 \alpha_{\pm} \equiv \frac{\tau_{R}}{2 \tau_{m}} (1 \pm \sqrt{1 - 2 i \omega \tau_{m}}) .
\end{equation}
The sums over $p$ in eq~\eqref{complex_modulus_rouse_exact} can be rewritten
by utilizing the partial fraction expansion for $\coth z$\cite{NIST-handbook}:
\begin{equation}
 \label{coth_partial_fraction_expansion}
 \coth z = \frac{1}{z} + 2 z \sum_{p = 1}^{\infty} \frac{1}{z^{2} + \pi^{2} p^{2}} .
\end{equation}
By combining eqs~\eqref{complex_modulus_rouse_exact} and \eqref{coth_partial_fraction_expansion},
we have
\begin{equation}
 \label{complex_modulus_rouse_exact_modified}
 \begin{split}
  G^{*}(\omega) 
  & = \nu k_{B} T 
  \frac{\alpha_{+}\alpha_{-}}{\alpha_{+} - \alpha_{-}}
  \left[ 
  \frac{\pi \sqrt{\alpha_{-}} \coth (\pi \sqrt{\alpha_{-}}) - 1}{2 \alpha_{-}}
  - \frac{\pi \sqrt{\alpha_{+}} \coth (\pi \sqrt{\alpha_{+}}) - 1}{2 \alpha_{+}}
  \right] \\
  & = \frac{\nu k_{B} T }{2}
  \left[ 
   \frac{\pi \alpha_{+}\alpha_{-}}{\alpha_{+} - \alpha_{-}}
  \left[ 
   \frac{\coth (\pi \sqrt{\alpha_{-}})}{ \sqrt{\alpha_{-}}}
  - \frac{\coth (\pi \sqrt{\alpha_{+}})}{\sqrt{\alpha_{+}}}
  \right] - 1
  \right] .
 \end{split}
\end{equation}

Eq~\eqref{complex_modulus_rouse_exact_modified} cannot be reduced to simpler
form in general.
At the intermediate angular frequency range where $\tau_{R}^{-1} \ll \omega \ll \tau_{m}^{-1}$,
we can approximate $\alpha_{\pm}$ as
\begin{equation}
 \alpha_{+} \approx \frac{\tau_{R}}{\tau_{m}} (1 - i \omega \tau_{m} / 2),
  \qquad
  \alpha_{-} \approx i \omega \tau_{R} / 2.
\end{equation}
$\cosh(z)$ can be approximated as $\cosh(z) \approx 1$ for $\mathrm{Re}(z) \gg 1$.
Then the complex modulus can be approximated as follows:
\begin{equation}
 \label{complex_modulus_rouse_exact_expanded}
 \begin{split}
  G^{*}(\omega) 
  & \approx \frac{\nu k_{B} T }{2}
  \left[ 
  \pi (i \omega \tau_{R} / 2)
  \left[ 
   \frac{1}{ \sqrt{i \omega \tau_{R} / 2}}
  - \frac{1}{\sqrt{(\tau_{R} / \tau_{m}) (1 - i \omega \tau_{m} / 2)}}
  \right] - 1
  \right] \\
  & \approx \frac{\pi \nu k_{B} T }{4}
  \left[ (1 + i)
  \sqrt{\omega \tau_{R}} - \left( \frac{\pi}{2}
    + i \omega \sqrt{\tau_{R}  \tau_{m}} \right)
  \right] .
 \end{split}
\end{equation}
The first term in the parenthesis in the last line of eq~\eqref{complex_modulus_rouse_exact_expanded}
corresponds to the complex modulus of the fully overdamped Rouse model.
Eq~\eqref{complex_modulus_rouse_exact_expanded} means that the complex
modulus is decreased (or the relaxation is accelerated) by the inertia effect, and this behavior is consistent
with one for the relaxation modulus.

Although eq~\eqref{complex_modulus_rouse_exact_modified}
is justified only for $\omega \tau_{m} \lesssim 1$, it would be informative
to see how it behave at the high frequency region ($\omega \tau_{m} \gg 1$).
$\alpha_{\pm}$ can be approximated as
\begin{equation}
 \alpha_{\pm} \approx \pm \frac{\tau_{R}}{2 \tau_{m}} \sqrt{- 2 i \omega \tau_{m}}
  = \pm \frac{\tau_{R}}{2 \tau_{m}} (1 - i)\sqrt{\omega \tau_{m}} .
\end{equation}
We can approximate eq~\eqref{complex_modulus_rouse_exact_modified} as
\begin{equation}
 \label{complex_modulus_rouse_exact_asymptotic}
 \begin{split}
  G^{*}(\omega) 
  & \approx \frac{\pi \nu k_{B} T }{2} \sqrt{\frac{\tau_{R}}{2 \tau_{m}}(1 - i) \sqrt{\omega \tau_{m}}} \\
  & \approx \frac{\pi \nu k_{B} T }{4} \sqrt{1 + 2^{-1/2}}
  [1 + (\sqrt{2} - 1) i] (\omega \tau_{R}^{2} / \tau_{m})^{1/4}.
 \end{split}
\end{equation}
Thus, for large $\omega$, the complex modulus depends on the angular
frequency as $G^{*}(\omega) \propto \omega^{1/4}$. (However, this behavior
cannot be observed, because we have momentum relaxation modes in such a
region.)

Secondly, we use the approximate expression for the relaxation modulus, eq~\eqref{relaxation_modulus_rouse_approx}.
The approximate expression for the complex modulus becomes
\begin{equation}
 \label{complex_modulus_rouse_approx}
 \begin{split}
  G^{*}(\omega) 
  & \approx
  \frac{ \nu k_{B} T}{4} \sqrt{\frac{\tau_{R}}{\tau_{m}}} 
  i \omega \int_{0}^{\infty} dt \, e^{ (1 / 4 \tau_{m} - i \omega) t}
  K_{1/4}
  \left( \frac{t}{4 \tau_{m}} \right)  \\
  & =
  \frac{ \nu k_{B} T}{4} \sqrt{\frac{\tau_{R}}{\tau_{m}}} 
  (1 - \beta) \int_{0}^{\infty} du \, e^{\beta u}
  K_{1/4}(u) ,
 \end{split}
\end{equation}
where we have defined $u \equiv t / 4 \tau_{m}$ and $\beta \equiv 1 - 0 - 4 i \omega \tau_{m}$ (we introduced $-0$ to
satisfy the condition $\mathrm{Re}(\beta) < 1$).
The integral over $u$ in eq~\eqref{complex_modulus_rouse_exact} can be simplyfied
by utilizing the following formula\cite{Bateman-integral-book} for $\mathrm{Re}(s) > -1$:
\begin{equation}
 \label{laplace_transform_modified_bessel}
 \int_{0}^{\infty} dt \, e^{- s t} K_{n}(t)
  = \frac{\pi \csc(n \pi)}{2} \frac{(s + \sqrt{s^2 - 1})^{n}
  - (s + \sqrt{s^2 - 1})^{-n}}{ \sqrt{s^2 - 1}}.
\end{equation}
From eqs~\eqref{complex_modulus_rouse_approx} and \eqref{laplace_transform_modified_bessel},
we have
\begin{equation}
 \label{complex_modulus_rouse_approx_modified}
 \begin{split}
  G^{*}(\omega) 
  & =
  \frac{\pi \nu k_{B} T}{4} \sqrt{\frac{\tau_{R}}{\tau_{m}}} 
  (1 - \beta) \frac{(\sqrt{\beta^2 - 1} - \beta )^{1/4}
  - (\sqrt{\beta^2 - 1} - \beta )^{-1/4}}{\sqrt{2} \sqrt{\beta^2 - 1}} \\
  & =
  \frac{\pi \nu k_{B} T}{4} \sqrt{\frac{\tau_{R}}{2 \tau_{m}}} 
  \frac{1 - \beta}{\sqrt{\beta^{2} - 1}}
  \left[ (\sqrt{\beta^2 - 1} - \beta)^{1/4}
  - (\sqrt{\beta^2 - 1} - \beta)^{-1/4} \right] .
 \end{split}
\end{equation}

Eqs~\eqref{complex_modulus_rouse_exact_modified} and
\eqref{complex_modulus_rouse_approx_modified} look rather different,
at least apparently. As before, we consider the case where $\tau_{R}^{-1} \ll \omega \ll \tau_{m}^{-1}$.
For $\omega \tau_{m} \ll 1$, we have
\begin{align}
 \frac{1 - \beta}{\sqrt{\beta^{2} - 1}}
 & \approx (- 1 + i) \sqrt{\omega \tau_{m}} , \\
  (\sqrt{\beta^{2} - 1} - \beta)^{1/4}
 & \approx \frac{1 - i}{\sqrt{2}} +
 \frac{i}{2} \sqrt{\omega \tau_{m}}  , \\
 (\sqrt{\beta^{2} - 1} - \beta)^{-1/4}
 & \approx \frac{1 + i}{\sqrt{2}}
 - \frac{1}{2} \sqrt{\omega \tau_{m}}  .
\end{align}
and we can approximate eq~\eqref{complex_modulus_rouse_approx_modified} as
\begin{equation}
 \label{complex_modulus_rouse_approx_expanded}
 \begin{split}
  G^{*}(\omega) 
  & \approx \frac{\pi \nu k_{B} T}{4} \sqrt{\frac{\tau_{R}}{2 \tau_{m}}} 
  (- 1 + i) \sqrt{\omega \tau_{m}}
  \left[ - \sqrt{2} i + \frac{- 1 + i}{\sqrt{2}} \sqrt{\omega \tau_{m}} \right] \\
  & \approx \frac{\pi \nu k_{B} T}{4}
  \left[  (1 + i) \sqrt{\omega \tau_{R}} - i \omega \sqrt{\tau_{R} \tau_{m}} \right] .
 \end{split}
\end{equation}
Eq~\eqref{complex_modulus_rouse_approx_expanded} has
the same form as eq~\eqref{complex_modulus_rouse_exact_expanded},
except an $\omega$-independent constant term.
For the case of $\omega \tau_{m} \gg 1$, we can approximate
\begin{equation}
 \frac{\beta - 1}{\sqrt{\beta^{2} - 1}} \approx -1,
\end{equation}
\begin{equation}
 \sqrt{\beta^{2} - 1} - \beta
  \approx - \frac{1}{2 \beta} \approx \frac{1}{8 i \omega \tau_{m}},
\end{equation}
and thus
\begin{equation}
 \begin{split}
 \label{complex_modulus_rouse_approx_asymptotic}
  G^{*}(\omega) 
  & \approx
  \frac{\pi \nu k_{B} T}{4} \sqrt{\frac{\tau_{R}}{2 \tau_{m}}} 
   (8 i \omega \tau_{m})^{1/4} \\
  & \approx
  \frac{\pi \nu k_{B} T}{4} 2^{-1/4}\sqrt{1 + 2^{-1/2}}
  [1 + (\sqrt{2} - 1)i] (\omega \tau_{R}^{2} / \tau_{m})^{1/4} .
 \end{split}
\end{equation}
Eq~\eqref{complex_modulus_rouse_approx_asymptotic} has the same angular
frequency dependence as eq~\eqref{complex_modulus_rouse_exact_asymptotic}
(although the absolute value is slightly different by the numerical factor $2^{-1/4} \approx 0.8409$.).
Therefore, we conclude
that the approximation employed to calculate eq~\eqref{relaxation_modulus_rouse_approx}
is reasonable.



\clearpage

\section*{Figure Captions}

Figure~\ref{dumbbell_model_relaxation_modulus}:
The shear relaxation modulus of a harmonic dumbbell with the inertia
effect. The momentum relaxation time is $\tau_{m} / \tau_{b} = 0.1$.
The dashed black curve is the exact result by
eq~\eqref{relaxation_modulus_harmonic_final}.
The solid black curve and the dotted gray curve are the approximations
for weak inertia (eq~\eqref{relaxation_modulus_harmonic_approx2_final})
and the fully overdamped limit (eq~\eqref{relaxation_modulus_harmonic_overdamped}), respectively.

\

Figure~\ref{dumbbell_model_viscosity_time}:
The zero shear viscosity $\eta_{0}$ and the longest relaxation time $\tau_{d}$
of a harmonic dumbbell with the inertia effect. Both $\eta_{0}$ and $\tau_{d}$
are normalized by their overdampled limits ($\nu k_{B} T \tau_{b} / 2$ and
$\tau_{b} / 2$, respectively).

\

Figure~\ref{rouse_model_relaxation_modulus}:
The shear relaxation modulus of the Rouse dumbbell with the inertia
effect, at the intermediate time region $\tau_{m} \lesssim t \lesssim \tau_{R}$. The momentum relaxation time is $\tau_{m} / \tau_{R} = 0.01$.
The solid black curve is the approximation
for weak inertia (eq~\eqref{relaxation_modulus_rouse_approx}),
and the dotted gray curve is modulus at the fully overdamped limit $G(t) = \nu
k_{B} T \sqrt{\pi \tau_{R} / 2 t}$. 
Black circles are results of the direct numerical calculations of
the discrete sum without the integral approximation (eq~\eqref{relaxation_modulus_rouse_exact}).
Gray diamonds are the results for the discrete sum at the fully
overdamped limit (eq~\eqref{relaxation_modulus_rouse_exact} with $\tau_{m}/\tau_{R} = 0$).


\section*{Figures}


\end{document}